# 10 Inventions on collapsible keyboards
## A TRIZ based analysis

**Umakant Mishra**
Bangalore, India
umakant@trizsite.tk
http://umakant.trizsite.tk

**Contents**

# 1. Introduction

A conventional computer keyboard consists of as many as 101 keys. The keyboard has several sections, such as text entry section, navigation section, numeric keypad etc. each having several keys on the keyboard. The keys are arranged quite spaciously so that the human fingers can operate comfortably. The number of keys and the arrangement of keys make a keyboard quite big.

Although a bigger keyboard is often comfortable to work with, they cannot be fit into laptop and small size computer boxes. The portable and handheld computers need small size keyboards. So there is a need to reduce the size of the keyboard to fit into the laptop box. There are various mechanisms to reduce the size of the keyboard; collapsible keyboard is one of them.

**Different mechanism of collapsible keyboards**

There are various types of collapsible keyboards disclosed in various inventions. Some of the interesting techniques are:

- Folding the keyboard by two fold to occupy less space.

- Folding the keyboard by threefold or multifold.

- Make the width of the keyboard compressible and expandable by using the space between the keys.

- Compress the height of the keys of the keyboard which preserves vertical space.

**Advantages of collapsible keyboards**

- Collapsible keyboards reduce the size of the keyboard so that they fit into the laptop box.

- Collapsible keyboards can reduce the size of the keyboard without needing to remove keys from the keyboard.

- Collapsible keyboards become bigger when expanded and gives typing comfort.

- Collapsible keyboards are expanded when used and compressed when not in use, thus giving the advantage of both a small and large keyboard.

- Collapsible keyboards are smaller when folded. They occupy less space and are easy to carry.

**Solving the contradiction of keyboard size and operational comfort**

Reducing the size of the keyboard usually leads to a contradiction. We need the laptop keyboard to be large for typing comfort. But we don't want the large keyboard to carry. We want the keyboard to be small to fit into the laptop box while carrying, but we want it to be large like a desktop keyboard when deployed. This creates a physical contradiction.

This contradiction is very effectively solved by a collapsible keyboard or folding keyboard. The folding keyboard provides all the benefits of a large keyboard and a small keyboard as well.

## 2. Inventions on collapsible keyboard

### 2.1 A tiltable keyboard for Portable computers (Patent 5168427)

**Background problem**

The keyboards on portable computers are flat and parallel to its base. When the laptop is on a table, the human arm is at an angle to the keyboard and not comfortable to operate a flat keyboard. For example when the laptop is on the table, a downwardly sloping orientation would make it convenient to operate.
Thus it is beneficial to have the ability of tilting the keyboard according the position of the laptop to the user.

**Solution provided by the invention**

Clancy et al. provided a solution to solve the above problem (Patent 5168427, assigned to Compaq Computer Corporation, Issued in Dec 1992). According to the invention, the keyboard structure is recessed within an open topped base housing portion of a compact notebook or laptop computer.
Front corner portions of the keyboard structure are secured to the base housing in a manner permitting the keyboard structure to be pivoted relative to the base housing.

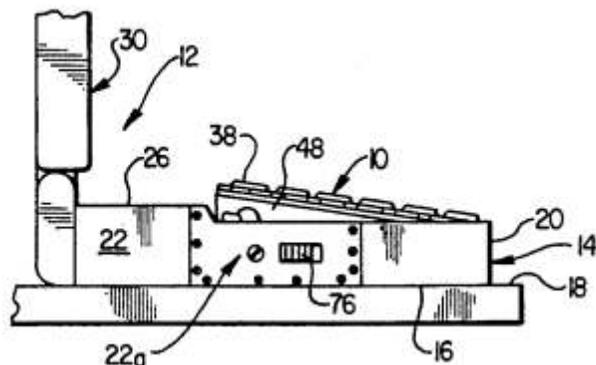

During storage or transport, the keyboard is locked with the housing in a parallel condition to the base housing. When the laptop is used, the keyboard lock can be opened to pop up the keyboard to an ergonomically improved tilted orientation. The structure of the keyboard slopes forwardly and downwardly for comfortable use. Although the keyboard can still be used in a locked flat position.

**TRIZ based analysis**

The keyboard should rest at the same angle as the operator's arm for comfortable operation **(desired result)**.

We want the keyboard of the laptop with a forwardly and downwardly sloping orientation, but we don't want to increase the overall exterior dimension of the laptop **(Contradiction)**.

The invention uses a tiltable keyboard, which can be sloppy while in use and flat when closed inside cover of the laptop **(Principle-15: Dynamize)**.

## 2.2 Sliding keyboard in a monolithic computer (Patent 5247428)

**Background problem**

A conventional computer is made of three main parts, the main box, a display unit and a keyboard. Traditionally they are provided separately for flexibility and other advantages, but that leads to the disadvantage of covering large desk space and increased cost of manufacturing and transportation.

**Solution provided by the invention**

Ching Yu invented a monolithic computer with a sliding keyboard (patent 5247428, Issued in Sep 93), which solves the space problem of the conventional computer. The invention gives a monolithic one-piece body, multifunctional computer having the peripheral extension interface, disk drive system and keyboard all assembled within the display unit.

According to the invention a hidden keyboard system is located in front of the display unit beneath the screen. The keyboard can be slide in and slide out as required. The disk drive system may be disposed at the right side of the display unit and accessed in a conventional way.

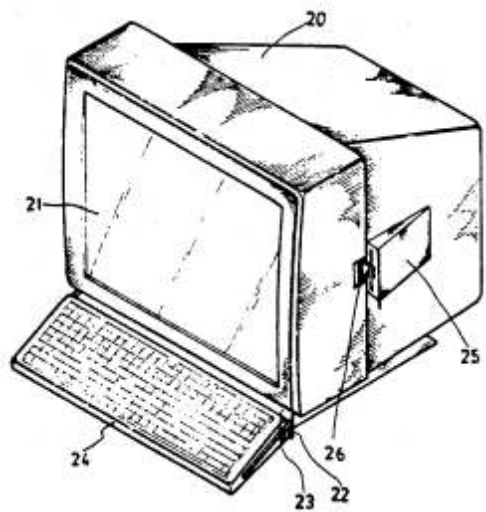

### TRIZ based analysis

The computer should be space efficient and cheap as well **(Desired result)**.
The conventional desktop is space hankering. The miniaturisation of a conventional desktop to a laptop makes it expensive **(Contradiction)**.

The invention makes use of the same components and devices of a conventional computer and reorganise its orientation to make it compact **(Principle–17: Another Dimension)**.

The keyboard is made sliding in structure, so that it can be pulled out when required for use and pushed in when not used **(Principle-15: Dynamize)**.

### 2.3 Folding and sliding keyboard for Personal Computer (Patent 5267127)

**Background problem**

The portable computers or laptops need small size keyboards that should fit into the laptop box. A typical problem in reducing the size of a keyboard is its inconvenience of operating compared to a standard sized keyboard. It is necessary to have a large keyboard while typing but a small keyboard while not using.

**Solution provided by the invention**

Pollitt invented a folding and sliding keyboard (Patent 5267127, assigned to IBM, Issued in Nov 93). The size of the keyboard is same as a standard keyboard when used, but very small (almost half) when dissociated in a stored position. The keyboard in the invention is segmented to two halves. The portions of the keyboard are designed for a pivotal movement on a keyboard axis.

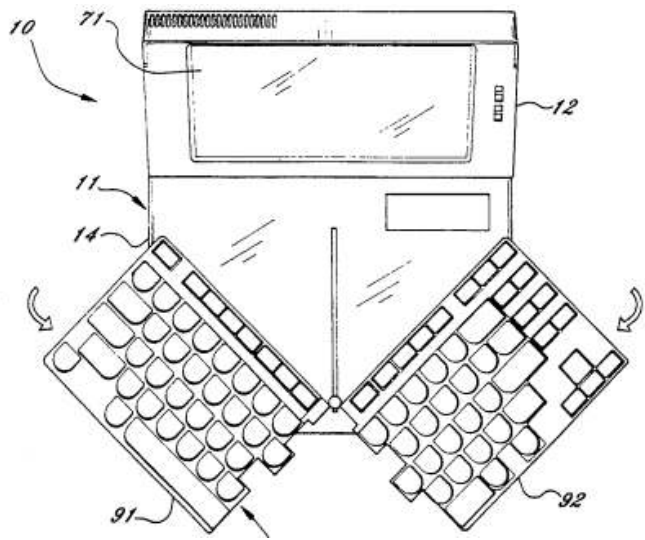

When the keyboard halves are slid to join together they form a complete keyboard of a standard size. When they are stored, they are slid back to have a dimension to fit within the enclosure.

**TRIZ based analysis**

The keyboard should change its size to big or small as and when required **(Ideal Final Result)**.
The size of a keyboard should be that of a standard desktop keyboard for typing comfort. But the size of the keyboard has to be small to fit into the laptop box **(Contradiction)**.

The invention splits the keyboard into two halves **(Principle-1: Segmentation)**.

The keyboard portions are designed for a pivotal movement, which facilitates folding and sliding **(Principle-15: Dynamize)**.

**2.4 Modular keyboard (Patent 5865546)**

**Background problem**
The modern keyboards are having more number of special purpose keys. The 83 keys of the early days' keyboards have been increased to 101 keys to include additional control and function keys. But this enhancement of the keyboard has certain drawbacks. (i) There are two sets of navigation keys out of which most users use only either of the sets. (ii) Besides the user has to use one or more external pointing devices each having a separate cable to connect to the rear side of the computer making the table messy. (iii) Moreover each device needs additional desk space.

**Solution provided by the invention**
Ganthier et al. found the solutions to all the above by disclosing a modular keyboard (US patent 5865546, Assigned to Compaq, Issued in Feb 99). The modular keyboard has several openings in which any of the input modules can be inserted. The user can insert a module and replace a module independently.

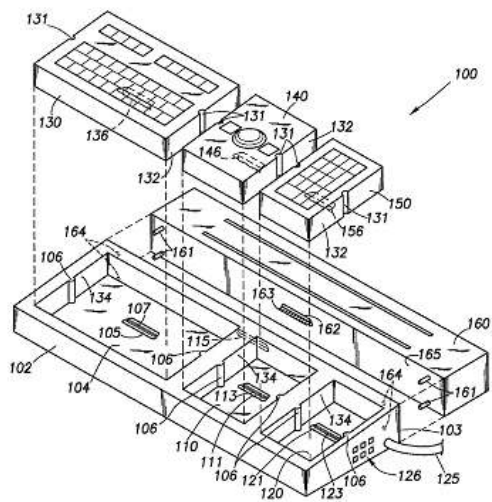

The disclosed modular keyboard has two additional slots other than the basic keyboard module. The user can use any two extra devices depending on need, either a numeric pad and a trackball, or a trackball and joystick or any two other devices depending on need. The user can plug in and plug out any device as required. A controller in the keyboard assembly will determine which types of input device modules are coupled with the keyboard assembly. This has the advantage of no extra cables, quick reconfiguration and saving of space.

**TRIZ based analysis**

Ideally the same keyboard should provide functions of a pointing device, navigation keys, joystick and other devices without needing extra cables and desk space for each device required **(Ideal Final Result)**.

The invention uses the same keyboard cable for all other external devices thus avoiding separate connections for each device to the computer **(Principle-6: Universality)**.

The invention proposes to connect upto two attachments such as pointing device, navigation keys or joystick etc. and remove the other attachments that are not used **(Principle-2: Taking out)**.

The invention allows connecting multiple pointing devices of our choice into some sockets on the keyboard **(Principle-7: Nesting)**.

We can select any two devices of our choice out of various devices including trackball, touchpad, joystick, pointing stick, or stylus etc. to plug into the keyboard **(Principle-15: Dynamize)**.

**2.5 Tri-fold personal computer with touchpad and keyboard (Patent 5926364)**

**Background**

The notebook computers are small, light but expensive. The desktop computers are powerful, cheaper but large. There is a need to combine the elements of both desktop and a notebook computer and get the benefits of both.

**Solution provided by the invention**

Karidis disclosed a solution (patent 5926364, assigned to IBM, July 99) of a hybrid packaging design for a portable personal computer. The invention comprises a tri-fold mechanical structure with a touch-screen display screen and a detachable keyboard, which is easy to pack with the computer. It has a stylus, a wireless remote control and other devices as well. The user can operate through the touchpad using fingers or a stylus.

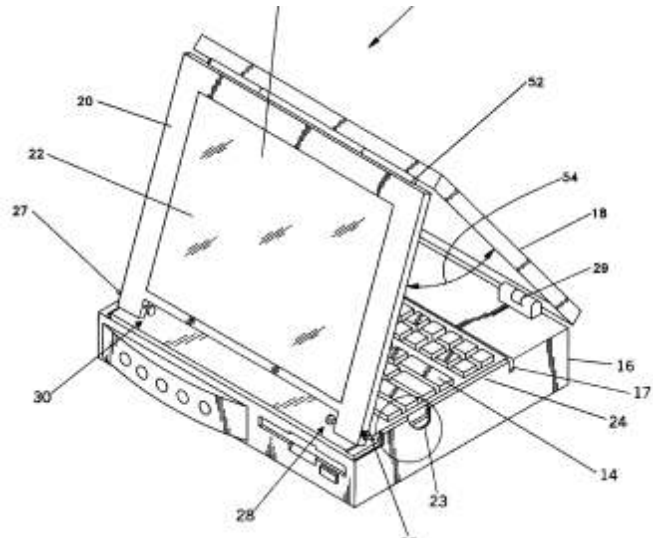

**TRIZ based analysis**

The small notebook computer should contain all the features of a desktop computer **(desired result)**.

Normally a notebook has a folded display unit **(Principle-15: Dynamize)**. The invention proposes to increase the number of folds to threefold **(Principle-17: Another dimension)**.

It uses a detachable keyboard **(Principle-15: Dynamize)**.

**2.6 Computer having a collapsible keyboard structure (Patent 5933320)**

**Background problem**

A small notebook keyboard is not comfortable for typing. On the other hand a big keyboard cannot fit in a laptop box. This leads to the classical contradiction of the keyboard size and operational comfort.

**Solution provided by the invention**

Malhi discloses a collapsible keyboard (patent 5933320, assigned to Texas Instruments, Aug 99) which can be used with notebooks. The keyboard can be placed in an expanded or compressed stage. While expanded, the size becomes the size of a standard keyboard for typing comfort. In compressed stage, the keyboard is folded on itself, typically when the computer is not in use to save the keyboard area.

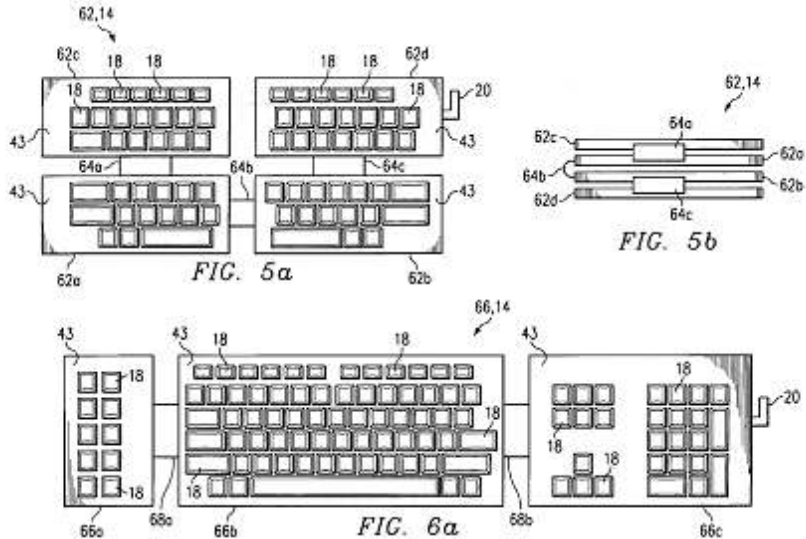

**TRIZ based solution**

The keyboard should be large for typing comfort, and small for carrying with laptop **(Contradiction)**.

The invention discloses a folding keyboard or collapsible keyboard **(principle-15: Dynamize)**.

The keyboard can be folded into threefold or even fourfold to save space **(Principle-17: Another dimension)**.

**2.7 Personal digital assistant having a foldable keyboard component (Patent 5941648)**

**Background problem**

The PDAs or organisers have very small space for the keyboard. This leads to the classical keyboard contradiction of keyboard size and operating comfort. Small keyboards are not convenient for typing like large keyboards. But there is a need for small keyboards, which are convenient for carrying.

**Solution provided by the invention**

Robinson, et al. invented a folding keyboard (Patent 5941648, assigned to Olivetti, Aug 99) for the small computers, PDAs and similar smaller devices.

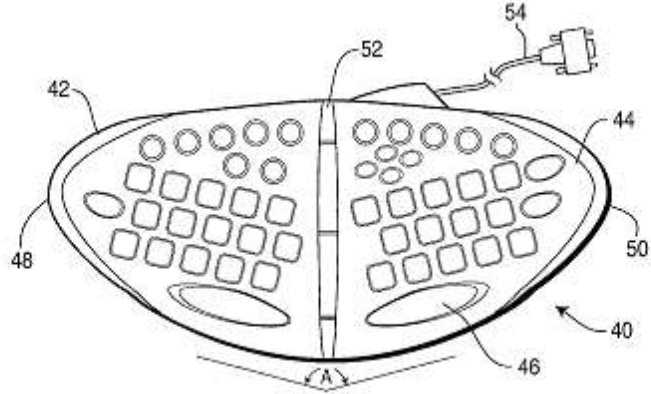

This invention provides a foldable keyboard which consists of two half keyboards consisting of half half keys complementing to each other. When in open position, the first half and the second half make a continuous keyboard. When the keyboard is closed, the first half will be folded over the second half.

**TRIZ based analysis**

The invention uses a folding mechanism to fold the keyboard while carrying **(Principle-15: Dynamize)**.

The keyboard halves have curved outer peripheral shapes **(Principle-14: Curvature)**.

**2.8 Portable computer having split keyboard and screen (Patent 5949643)**

**Background problem**

The classical problem exists as usual with the size of the keyboard for a portable computer. The small computing devices like PDA and notebook do not have space to accommodate large keyboards. But small keyboards are neither efficient nor comfortable for typing. Although there are several inventions attempting to solve the same problem, it is still required to find a better solution.

**Solution provided by the invention**

Batio invented a folding keyboard (US patent 5949643, Sep 99) for the PDA, notebook and other small computing devices. The keyboard consists of two, pivotally hinged halves, which make a perfectly flat horizontal keyboard. Each half has its own set of keys and space bar. The keyboard also has a pointing device and a joystick adapter. The invention also provides a dual split screen where each half of the split screen is pivotally mounted for universal rotation.

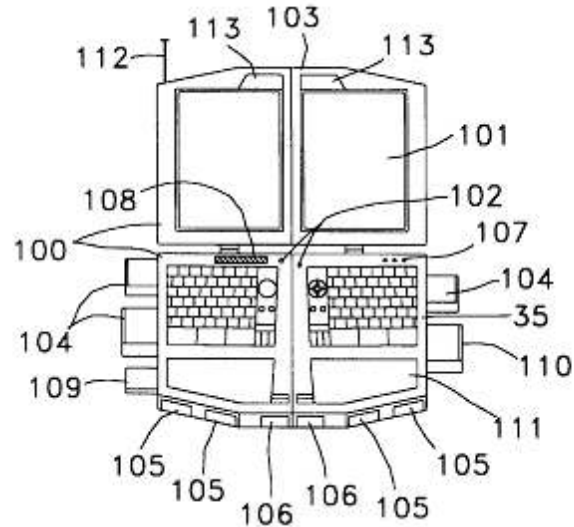

This invention is not confined to only the folding nature of the keyboard. It also has two independent dedicated microprocessors. While one microprocessor use used as a notebook computer, the other microprocessor may be dedicated for video games or a set-top converter box for Internet access through cable TV.

**TRIZ based analysis**

The invention splits the keyboard into two halves. It also splits the screen into two halves **(Principle-1: Segmentation)**.

The two halves of the keyboard and screen have folding joints, which make it possible to open during usage and fold during storage **(Principle-15: Dynamize)**.

**2.9 Keyboard for a handheld computers (Patent 6507336)**

**Background problem**

The personal digital assistant (PDA) and palm-sized computers also need input devices like keyboards, stylus or touch sensitive displays. However, a stylus is more efficient for graphics input and not for entering text. Keyboards (even folding keyboards) are too large and cumbersome for handheld computers as the complete device is intended to be kept inside a pocket. Secondly a palm top is usually held with one hand and operated with another; hence an add-on keyboard is not suitable to operate.

**Solution provided by the invention**

Lunsford developed a new keyboard (patent 6507336, assigned to Palm Inc., Jan 03) to overcome the difficulties of separate physical keyboard. The invention discloses a keyboard actuated by a stylus of the handheld computer.

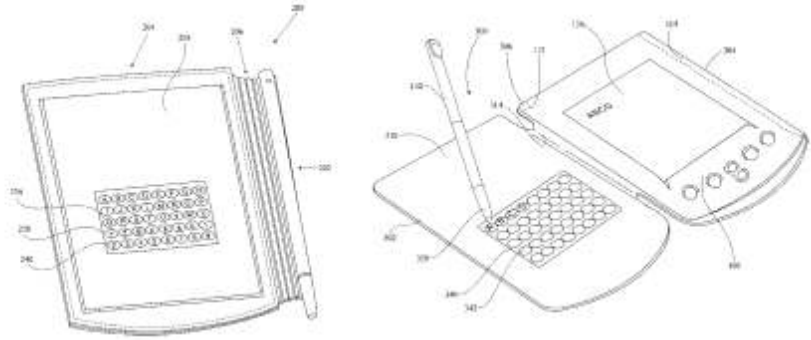

This implementation is better than a folding keyboard as it is more compact and attached to the computer. It is better than a soft keyboard or virtual keyboard displayed on the screen. A stylus-based keyboard is also convenient for physically handicapped users who can hold the stylus on mouth and operate the keyboard.

While the keys in the typical keyboard are convex for easy access by human fingers, the keys in the invention are concave for easy access by the stylus. The concave surface is kept below the surface of the encasement cover, which avoids undesirable contact or friction against the display screen.

**TRIZ based analysis**

The invented keyboard is actuated with a stylus instead of human fingers **(Principle-28: Mechanics Substitution)**.

Contrary to the conventional convex keys of a typical keyboard, the invention uses concave keys for easy access by a stylus **(Principle-13: Other way round)**.

**2.10 Portable electronic device with a concealable keyboard module (Patent 6556430)**

**Background problem**

The conventional model of notebook computers includes a keyboard and a touch pad, which covers all the space on the surface and leaves no space to install other controls. It is necessary to create space on the laptop surface to install other controls like touch pad, pen control etc.

**Solution provided by the invention**

Kuo et al. invented a concealed keyboard (Patent Number: 6556430, assigned to Compal electronics, Apr 03) which can be used with a notebook or PDA or similar small devices. The keyboard remains inside the main module and can be pulled out via an opening. This mechanism leaves enough space for the touch control, pen control and other devices to be fixed on the main surface of the laptop.

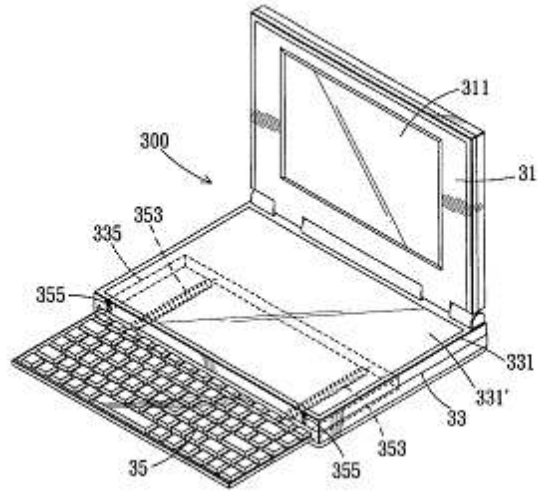

**TRIZ based analysis**

The invention makes a drawer type keyboard concealed within the laptop, which can be pulled or pushed easily according to usage **(Principle-34: Discard and Recover)**.

The surface of the keyboard is used for installing other controlling devices **(Principle- Another Dimension)**.

## 3. Summary and conclusion

Although different inventions intend to achieve the same objective of reducing the keyboard size, they all differ in their mechanism. For example, some invention uses a compression mechanism, some invention uses a folding mechanism, and some invention uses a collapsing mechanism and so on.

Multi-folding keyboards have not been very popular because of their delicacy and high cost. But single folding and tilting keyboards have been quite well adopted.

As there is a genuine demand for small size keyboards there will be many more inventions in the same field in future days. All the above inventions have tried to solve the classical contradiction of keyboard size and operating comfort. We can expect to see many more inventions on this technology in future to see collapsible keyboards, which is both comfortable for typing and easy for carrying.

3. 5267127, "Folding and sliding keyboard for personal computer, inventor Richard Pollitt, assigned to IBM, Nov 93

4. 5865546, "Modular keyboard", Ganthier et al., Assigned to Compaq, Feb 99

5. 5926364, "Tri-fold personal computer with touchpad and keyboard", John P. Karidis, assigned to IBM, Jul 99

6. 5933320, "Computer having a collapsible keyboard structure", Satwinder Malhi, assigned to Texas Instruments, Aug 99

7. 5941648, "Personal digital assistant having a foldable keyboard component", Robinson, et al., assigned to Olivetti, Aug 99

8. 5949643, "Portable computer having split keyboard and screen", Jeffry Batio, Sep 99

9. 6507336, "Keyboard for handheld computers", Lunsford, assigned to Palm Inc., Jan 03

10. 6556430, "Portable electronic device with a concealable keyboard module", Kuo et al., assigned to Compal electronics, Apr 03

11. US Patent and Trademark Office (USPTO) site, http://www.uspto.gov/

**10 Inventions on collapsible keyboards, by Umakant Mishra**　　　　　　　　　　http://www.trizsite.tk